\documentclass[aps,prd,showpacs,preprintnumbers,amsmath,amssymb,
               nofootinbib,superscriptaddress]{revtex4}
\usepackage{graphicx}
\usepackage{verbatim}
\usepackage{slashed}
\usepackage{amsthm}
\usepackage{amsfonts}
\usepackage[dvips]{color}

\def\beqn#1{\begin{equation}\label{#1}}
\def\eeqn{\end{equation}}

\def\oneR{\ensuremath{\underline{1}}}
\def\twoR{\ensuremath{\underline{2}}}
\def\thrR{\ensuremath{\underline{3}}}
\def\fouR{\ensuremath{\underline{4}}}
\def\fivR{\ensuremath{\underline{5}}}
\def\sixR{\ensuremath{\underline{6}}}
\def\sevR{\ensuremath{\underline{7}}}
\def\ninR{\ensuremath{\underline{9}}}
\def\eleR{\ensuremath{\underline{11}}}

\def\ththMat#1#2#3#4#5#6#7#8#9{\ensuremath{\begin{pmatrix}#1&#2&#3\\
                                                          #4&#5&#6\\
                                                          #7&#8&#9\end{pmatrix}}}
\def\twoneMat#1#2{\ensuremath{\begin{pmatrix}#1\\
                                              #2\end{pmatrix}}}
\def\twtwMat#1#2#3#4{\ensuremath{\begin{pmatrix}#1&#2\\
                                                #3&#4\end{pmatrix}}}

\def\order#1{\ensuremath{\mathcal{O}\left(#1\right)}}

\begin{document}

\title{Exotic fermion multiplets as a solution to baryon asymmetry, dark matter and neutrino masses}

\author{Chuan-Hung~Chen}
\email{physchen@mail.ncku.edu.tw}
\affiliation{Department of Physics, National Cheng-Kung University, Tainan 701, Taiwan}
\affiliation{National Center for Theoretical Sciences, Hsinchu 300, Taiwan}
\author{Sandy~S.~C.~Law}
\email{slaw@mail.ncku.edu.tw}
\affiliation{Department of Physics, National Cheng-Kung University, Tainan 701, Taiwan}

\date{15 February, 2011}

\begin{abstract}
We propose an extension to the standard model where three exotic fermion 5-plets and one scalar 6-plet are added to the particle content. By demanding that all interactions are renormalizable and standard model gauge invariant, we show that the lightest exotic particle in this model can be a dark matter candidate as long as the new 6-plet scalar does not develop a nonzero vacuum expectation value. Furthermore, light neutrino masses are generated radiatively at one-loop while the baryon asymmetry is produced by the CP-violating decays of the second lightest exotic particle. We have demonstrated using concrete examples that there is a parameter space where a consistent solution to the problems of baryon asymmetry, dark matter and neutrino masses can be obtained.
\end{abstract}

\maketitle


\section{Introduction}

The observation of neutrino oscillations \cite{nu_exp}, which provides compelling evidence for nonzero neutrino masses, has long indicated to us that the minimal standard model (SM) is incomplete and must be extended. Meanwhile, astrophysical and cosmological data have given us ample important hints regarding the missing mass of the universe \cite{Nakamura:2010zzi,Bertone:2004pz}, prompting many theorists to build new models for accommodating such dark matter (DM). So, there is no shortage of motivation for developing physics beyond the SM.

In addition to these, it is well-known that the problem of generating a primordial baryon asymmetry \cite{Komatsu:2010fb,Nakamura:2010zzi} must also bring together the studies of particle physics and cosmology. Most notably, in models of thermal leptogenesis \cite{Fukugita:1986hr}, the SM is extended to include lepton violating interactions so that baryogenesis can be originated from a lepton asymmetry while neutrino masses are naturally generated via the type-I seesaw mechanism \cite{type1seesaw}. In the light of this, it is perhaps most interesting to explore extensions to the SM that can address all three issues of baryon asymmetry, dark matter and neutrino masses simultaneously.

To construct a model that can achieve this, we first note that a popular candidate for the role of the non-baryonic DM is the weakly interacting massive particle (WIMP) \cite{Bertone:2004pz}. A key feature in the WIMP-like models is that they typically require an extra symmetry beyond the SM to ensure the stability of the DM candidate (e.g. R-parity in supersymmetric theories). However, such new symmetries are often introduced solely for the purpose of stabilizing the DM and nothing else, making them rather ad-hoc. 
Thus, it may be better to build up from a ``minimal DM'' approach \cite{MDM} where renormalizability and SM gauge invariance are the only requirements. It has already been shown that realistic DM models based on this idea can be built \cite{MDM, XGMDM}, and they involve exotic electroweak multiplets of particles.

Secondly, it was pointed out in the seminal work of \cite{Ma:2006km} that there could be a natural link between WIMP DM and neutrinos if the neutrino masses are generated radiatively. Such link is possible when there is a symmetry which forbids any particles except the left-handed (LH) lepton doublet to couple to the DM candidate, and the Higgs potential is arranged such that this symmetry remains unbroken after spontaneous symmetry breaking.

Our primary aim in this work is to present an exotic multiplet model which can be a consistent solution to neutrinos, DM and baryon asymmetry by amalgamating the two observations above. We shall demonstrate in the subsequent sections that by introducing a set of exotic multiplets with the appropriate SM transformation properties, the lightest exotic particle can be the DM candidate; neutrinos can develop a mass via one-loop diagrams; while baryogenesis can be achieved via the CP asymmetric decays of the second lightest exotic particles.

\section{The model}\label{sec:model}

In order to construct a consistent solution to baryon asymmetry, dark matter and neutrino masses, we extend the particle content of the minimal SM by adding \emph{three} right-handed (RH) fermion multiplets, $N_k, (k =1,2,3)$, and \emph{one} scalar multiplet, $\chi$, with SM gauge transformation properties:
\beqn{eqn:5_6plets}
 N_k  \sim (1,5,0), \text{ for all } k, \quad\text{and}\quad
 \chi \sim (1,6,-1/2)
\eeqn
respectively. This is in fact the most minimalist choice one may choose for $N_k$ and $\chi$ if one adopts the ``minimal DM'' idea of \cite{MDM} and demands that the lightest fermion 5-plet here be the dark matter. Other smaller multiplet combinations for these new particles will result in unwanted terms in the scalar potential which can destabilize our dark matter candidate, or give phenomenologically unacceptable component fields that have fractional charges (see \cite{XGMDM} for a good discussion on this point). The main reason for introducing three (rather than a lesser number of) $N_k$'s is because it is the smallest number such that all allowable light neutrino mass patterns can be accommodated by the model.  With the inclusion of the new fields, the Lagrangian of interest is 
\beqn{eqn:main_Lag}
 \mathcal{L}_\text{int} =
  i \overline{N}_k \slashed{D} N_k
  + (D^\mu\chi)^\dagger (D_\mu\chi)
  -\left[ h_{jk}\, \overline{L}_j \,\chi\,N_k 
          + \frac{1}{2} \overline{(N_k)^c}\,M_k\,N_k + h.c.
        \right]
  - V_S\;,
\eeqn
where $L_j = (\nu,\ell)_j^T$ is the $j$-flavor LH lepton doublet, $D_\mu$ denoting the SM covariant derivative, and $h_{jk}$ is the $(j,k)$-element of the Yukawa coupling matrix which is assumed to be complex. Without loss of generality, we have chosen to work in the basis where $M_k$ (the $N_k$ mass matrix) is real and diagonal.
The scalar potential, $V_S$, is given by
\beqn{eqn:V_S}
 V_S = \mu_\phi^2 \phi^\dagger\phi
       + \mu_\chi^2 \chi^\dagger \chi
       +\frac{\lambda_\phi}{2}\left(\phi^\dagger\phi\right)^2
       +\frac{\lambda_{\chi\alpha}}{2}\left(\chi^\dagger\chi\right)^2_\alpha
       +\lambda_{\phi\chi\beta} \left(\phi^\dagger\phi
          \chi^\dagger\chi
            \right)_\beta
       +\frac{1}{2}\left[ {\lambda}_{\phi\chi}' (\phi\chi)^2 + h.c.
         \right]\,,     
\eeqn
with $\phi = (\phi^+, \phi^0)^T$ being the SM Higgs doublet which transform as $(1,2,1/2)$ under the SM gauge group. To simplify the subsequent discussion on how potential $V_S$ can ensure that there is a dark matter candidate in this model and the stability of the vacuum solution, we will assume that all couplings in (\ref{eqn:V_S}) are real. The subscripts $\alpha$ and $\beta$ in (\ref{eqn:V_S}) denote the many independent ways to contract the components of the SU(2) multiplets involved, and we sum over them. For example, the term 
$\phi^\dagger\phi\chi^\dagger\chi$, which is a tensor product of representations $\twoR^* \otimes\, \twoR\, \otimes\, \sixR^* \otimes\, \sixR$ can be contracted in the following ways (see Appendix~\ref{App:CG_coef} for the relevant Clebsch-Gordon coefficients)
\beqn{eqn:2266contract}
 \underbrace{\twoR^* \otimes\;\twoR}_{\oneR + \thrR} \;\otimes\;
 \underbrace{\sixR^* \otimes\;\sixR}_{\oneR + \thrR + \cdots}
 \quad\text{or}\quad
 \underbrace{\twoR \;\otimes\;\sixR}_{\fivR + \sevR} \;\otimes\;
 \underbrace{\twoR^* \otimes\;\sixR^*}_{\fivR^* + \sevR^*}
  \quad\text{or}\quad
 \underbrace{\twoR \;\otimes\;\sixR^*}_{\fivR^* + \sevR^*} \;\otimes\;
 \underbrace{\twoR^* \otimes\;\sixR}_{\fivR + \sevR} \;.
\eeqn
Note, however, that not all possible expansions in (\ref{eqn:2266contract}) are independent of each other, in the sense that some of the resulting singlets are actually linear combinations of other possible singlets.
Similar conclusions can also be drawn for the contraction of the other terms in $V_S$.

A careful analysis of the Lagrangian in (\ref{eqn:main_Lag}) will reveal that the lightest neutral component of the newly introduced multiplets can be a dark matter candidate as long as the parameters of $V_S$ are such that scalar 6-plet $\chi$ does \emph{not} develop a nonzero vacuum expectation value (VEV). This is because when $\langle\chi\rangle =0$, the lightest exotic particle has no means to decay in this model.\footnote{When $\langle\chi\rangle \neq 0$, cross terms such as $\langle\chi\rangle \chi\phi\phi$ will allow the lightest exotic particle to decay into SM particles.} For the $V_S$ depicted, this vacuum condition can be achieved by choosing all couplings which involve $\chi$ to be positive. It will become apparent later that such freedom to tune the parameters will not be affected by any of the constraints coming from successful baryogenesis, dark matter or neutrino masses. 

We would like to emphasize that this model gives rise to a dark matter candidate, not because we have imposed a new symmetry (local or global) by hand which forbids their decays, but because the Lagrangian in (\ref{eqn:main_Lag}) possesses an accidental $Z_2$ symmetry ($N_k \rightarrow -N_k$ and $\chi \rightarrow -\chi$ while $\psi_\text{SM} \rightarrow \psi_\text{SM}$). This $Z_2$ symmetry will remain unbroken when $\langle\chi\rangle =0$ is the vacuum solution. 
One might worry that higher loop corrections will introduce possible instabilities for the vacuum $\langle\chi\rangle =0$ at very high energies when the ``negative'' running of the couplings changes the profile of potential $V_S$. While this is a valid concern given that the term, $\overline{L}_j \,\chi\, N_k$ in (\ref{eqn:main_Lag}) can lead to fermionic loops running in the higher order diagrams for the quartic couplings, the situation can be placed under control with careful choices for the tree-level couplings $\lambda_{\chi\alpha}$ and $\lambda_{\phi\chi\beta}$ for a given cut-off scale. 
In addition, observe that the quartic terms involving ${\lambda}_{\phi\chi}'$ do not get such ``negative'' running at the one-loop level. Thus, to a good approximation, we can assume ${\lambda}_{\phi\chi}'$ to be unrestricted by the vacuum stability conditions.

With this setup, either $N_k$ or $\chi$ can play the role of the dark matter depending on which one has the smallest mass. Our model building choice here is to make the lightest\footnote{We will use the convention $M_1<M_2<M_3$ throughout the paper. Whether or not this hierarchy should be strong or quasi-degenerate shall be discussed in later sections.} 5-plet Majorana fermion, $N_1$, to be our dark matter candidate, and hence, we demand that $M_1 < M_\chi$. It should be pointed out that, in reality, only the neutral component, $N^0_1$, of the 5-plet can be the dark matter. This comes about because one-loop electroweak corrections will introduce a mass splitting between the different components of the multiplet which are otherwise mass degenerate at tree-level. The mass difference induced by SM gauge boson loops for two different components of $N_k = (N_k^{++},N_k^{+},N_k^{0},N_k^{-},N_k^{--})^T$ having electric charges $Q_1$ and $Q_2$ is given by \cite{MDM}
\beqn{eqn:mass_split}
 M_k^{Q_1} - M_k^{Q_2}
 =\frac{g^2 M_k}{16\pi^2} \left\{
     \sin^2\theta_w\left(Q_1^2-Q_2^2\right) 
       \widetilde{f}\left(\frac{M_Z}{M_k}\right) + (Q_1+Q_2-2Y)(Q_1-Q_2)
       \left[ \widetilde{f}\left(\frac{M_W}{M_k}\right)-
        \widetilde{f}\left(\frac{M_Z}{M_k}\right) \right]
 \right\}\;,
\eeqn
where hypercharge $Y=0$ for our 5-plet, $\theta_w$ is the Weinberg angle and
\beqn{eqn:mass_split_f}
 \widetilde{f}(x) = \frac{x}{2}\left[
  3x^3 \ln x - 2x + (x^2+2)\sqrt{x^2-4}\;
   \ln \left(\frac{x^2-2-x\sqrt{x^2-4}}{2}\right)
 \right]\;.
\eeqn
Therefore, in the relevant limit\footnote{We shall show later that, for consistency, $M_k$ will be at least $\mathcal{O}(10)$~TeV.} of $M_k\gg M_{W,Z}$, the loop function $\widetilde{f}(x\ll 1) \simeq 2\pi x + \mathcal{O}(x^2)$, and the mass difference is approximately
\beqn{eqn:mass_split_appx}
 M_k^{Q_1} - M_k^{Q_2} \simeq \frac{g^2 M_W}{4\pi}
  \left(Q_1^2-Q_2^2\right) \sin^2\left(\frac{\theta_w}{2}\right)\;.
\eeqn
This means that the component $N^{\pm}_k$  is about 166~MeV heavier than the $N^0_k$. As a consequence, all the charged components of $N_k$ are unstable since processes like $N^{\pm}_k \rightarrow N^0_k \pi^{\pm}$ are kinematically accessible. Similar mass splittings occur for the 6-plet scalar $\chi$. But since we have already assumed $M_1 < M_\chi$, the exact mass spectrum of all its components will be inconsequential to our analysis here. 
For completeness though, and to display the connection to the parameters in (\ref{eqn:V_S}), we have listed the tree-level masses for the $\chi$ component fields after spontaneous symmetry breaking ($\langle\phi\rangle = (0,u)^T$ and $u \neq 0$) in Appendix~\ref{app:chi_mass}.

In the following sections, we will demonstrate that there is a consistent parameter space where this model of exotic multiplets can solve the problems of baryogenesis, dark matter and neutrino masses.

\section{The dark matter candidate}\label{sec:DM}

In order to perform the analyzes for baryogenesis and neutrino masses later on, we must first compute the relevant parameters for our dark matter candidate, $N_1^0$. In particular, we need to understand how the mass scale of $N_1$ (and consequently, the lower bound on all other exotic particles in the model) is constrained by cosmology.

Like other WIMP models, the ability to reproduce the observed DM relic density is a crucial ingredient in determining whether the DM candidate under investigation is a viable one. Assuming the standard thermal freeze-out of $N_1$ is solely responsible for the DM relic that we measure,
$\Omega_\text{CDM} h^2 = 0.110 \pm 0.006$ \cite{Komatsu:2010fb, Nakamura:2010zzi, Bertone:2004pz},\footnote{Here, $h$ denotes the Hubble constant in units of 100 km $\rm{s}^{-1} \rm{Mpc}^{-1}$. It is not related to the Yukawa coupling matrix, $h$ in (\ref{eqn:main_Lag}).} connection between the $N_1$ annihilation rate and $\Omega_\text{CDM} h^2$ will then fix the scale for $M_1$. The typical condition that one employs is \cite{Bertone:2004pz}:
\beqn{eqn:relic_rule}
 \Omega_\text{CDM} h^2 \simeq 
 \frac{3\times 10^{-27} \text{ cm}^3 \text{s}^{-1}}{\langle \sigma_A v\rangle}\;,
\eeqn
where $\sigma_A$ is the total annihilation cross-section, $v$ is the relative velocity between the annihilating pair, and $\langle ...\rangle$ denotes thermal averaging. In the model we are considering, $\sigma_A$ is the cross-section for annihilating a component of $N_1$ with another suitable counterpart of the multiplet into SM particles\footnote{In the hot early universe (and given the smallness of the mass splitting), all components of $N_1$ may be treated as degenerate, and therefore, charged components of $N_1$ should be included in the analysis here. However, for consistency with later sections, $N_{2,3}$ is not included because we shall assume that $M_1 \ll M_{2,3}$.}. Furthermore, we will assume that these coannihilations happen predominantly via interactions mediated by SM gauge bosons. 
   
The canonical method to compute the DM relic abundance is to solve the Boltzmann evolution equation using the relevant annihilation cross-section. In the notations of our model, we have
\beqn{eqn:BE_DM}
 \frac{dY_1}{dz'} = -\frac{s}{H z'} \left(Y_1^2 - (Y_1^\text{eq})^2 \right) \langle\sigma_A v\rangle \;,
 \qquad z'\equiv \frac{M_1}{T} \;,
\eeqn
where $Y_1$ denotes the $N_1$ number density per entropy, $s$, per comoving volume (with $Y_1^\text{eq}$ being the corresponding density at thermal equilibrium), $H$ is the Hubble parameter and $T$ is the temperature of the universe. The elements which govern our cross-section $\langle\sigma_A v\rangle$ originate from the first kinetic term in (\ref{eqn:main_Lag}), and it may be expanded in component form as
\begin{align}
 \sqrt{10}\, \mathcal{L}_\text{kin} &=
  \sqrt{2} (g\cos\theta_w Z_\mu - e A_\mu)
   \left( 2\overline{N_1^{++}}\gamma^\mu N_1^{++} 
          + \overline{N_1^{+}}\gamma^\mu N_1^{+}
          - \overline{N_1^{-}}\gamma^\mu N_1^{-}
          - 2\overline{N_1^{--}}\gamma^\mu N_1^{--}
    \right)
   \nonumber\\
   &\quad +g\left[ W_\mu^+ \left(
     \overline{N_1^{++}}\gamma^\mu N_1^{+}
    +\overline{N_1^{+}}\gamma^\mu N_1^{0}
    +\overline{N_1^{0}}\gamma^\mu N_1^{-}
    +\overline{N_1^{-}}\gamma^\mu N_1^{--}
   \right)
   + h.c.
   \right]
     \;,\quad e > 0\;.
\end{align}
Through these terms, components of $N_1$ can coannihilate into a pair of SM particles either via a $t$- or $u$-channel $N_k$ or a $s$-channel gauge boson. The computation of the total cross-section was performed in full in the last reference of \cite{MDM}. It was found that (in the non-relativistic limit, i.e. $v\ll 1$):
\beqn{eqn:sigmaA_v}
 \langle \sigma_A v \rangle \simeq
  \frac{1}{2 g_N^2 M_1^2} \left[ \left(1+ 
      \frac{\langle v^2\rangle}{4}\right) c_s 
  +\frac{\langle v^2\rangle}{2}\, c_p\right]\;,
\eeqn
where $g_N = 2\times 5 =10$ is the degrees of freedom for Majorana 5-plet $N_1$. The $s$- and $p$-wave coefficients are respectively
\beqn{eqn:s_p_coeff}
 c_s = \frac{1035 \,g^4}{8\pi}\;,
 \qquad
 c_p = \frac{1215 \,g^4}{8\pi}\;.
\eeqn
The Boltzmann equation (\ref{eqn:BE_DM}) may now be solved, taking into account the non-perturbative corrections from electroweak Sommerfeld enhancement \cite{sommerfeld}. When the dust has settled, it is determined that the observed DM relic abundance can be explained if \cite{MDM}
\beqn{eqn:M_1_mass}
 M_1 = 9.6 \pm 0.2 \text{ TeV}\;.
\eeqn
This result is in line with the typical WIMP type models where TeV scale DM are predicted. The fact that the limit on $M_1$ turns out to be slightly bigger than the usual WIMP reflects the extra coannihilation contributions the 5-plet $N_1$ (with almost degenerate components) is providing\footnote{Sommerfeld effects have increased the final result in (\ref{eqn:M_1_mass}) by a factor of about 2 as well \cite{MDM}.}. Note that any remaining charged components of $N_1$ will eventually decay away before nucleosynthesis (due to their slightly higher mass) with negligible entropy release.

It is worth mentioning that although a DM mass of the size shown in (\ref{eqn:M_1_mass}) cannot be tested at the LHC, it has been shown in \cite{MDM} that such 5-plet DM may be within the reach of future 
DM direct detection experiments like SuperCDMS  \cite{SuperCDMS} and XENON-1T \cite{XENON1T}.

\section{Baryon asymmetry via Leptogenesis}\label{sec:lepto}

While baryogenesis remains an open problem with many candidate solutions, one popular way to tackle it is via leptogenesis, whereby a lepton asymmetry, $\Delta_{L}$, is first created by some $L$-violating processes operating in the early universe, and then partially converted to the required cosmic baryon asymmetry by non-perturbative electroweak sphaleron interactions. 

The classic scenario of leptogenesis \cite{Fukugita:1986hr} (and many subsequent extensions) is based upon the type-I seesaw \cite{type1seesaw} Lagrangian where (at least two) RH electroweak singlet neutrinos, $\nu_R$, are introduced to the lepton sector with Yukawa couplings, $\overline{L}\, \phi\, \nu_R$. As a result, the heavy Majorana $\nu_R$ can decay into $L \phi^\dagger$ during the primordial times and give rise to an excess in $\Delta_{L}$. A major advantage of this setup over many other baryogenesis models is that the type-I seesaw Lagrangian automatically solves the light neutrino mass problem, providing a link (albeit indirect) between the parameters of neutrino physics and cosmology. Other variations to the general scheme exist\footnote{See \cite{lepto_examples} for an incomplete list of examples.}, and they provide interesting alternative solutions that can lead to other implications.\footnote{In fact, the model presented in this paper is one such variation.}

Over the years, the sophistication in the quantitative analysis of leptogenesis has  improved dramatically. Most notably, the careful treatment of the different washout \cite{listwashout} and its generalization to include flavor effects \cite{Vives:2005ra,flavor_eff, flavor_eff_Nardi, JosseMichaux:2007zj, Antusch:2010ms, Bertuzzo:2010et} have been pivotal to the understanding of the workable parameter space for many interesting leptogenesis scenarios. One such scenario that is particularly relevant to our discussion here is the so-called ``$N_2$-leptogenesis'' \cite{DiBari:2005st, Bertuzzo:2010et, Antusch:2010ms}, where the lepton asymmetry is predominantly produced by  the decays of the next-to-lightest RH neutrinos (rather than by the lightest one as in the conventional setup). We shall show that there is a strong resemblance between the model presented in this paper and the typical ``$N_2$-leptogenesis'' setup. As a result, the constraints on the leptogenesis parameter space for them will be markedly similar, which will ultimately allow us to construct an existence proof of successful leptogenesis for our exotic multiplet model.

\subsection{Our model in detail}

To begin with, let us recall that we have $M_\chi > M_1 \simeq 10$~TeV being established from the dark matter constraints in the previous section.
Therefore, in order for our 5-plet fermion, $N_2$ to be eligible for ``$N_2$-leptogenesis'', we must now further assume that
\beqn{eqn:full_mass_rel}
 M_3 \gg M_2 > M_\chi > M_1 \simeq 10 \text{ TeV}\;.
\eeqn
Hence, it will be natural for the decay, $N_2\rightarrow L_j \chi^\dagger$, to proceed via the Yukawa term in Lagrangian (\ref{eqn:main_Lag}). Note that at this temperature scale, the mass splittings of the component fields in the multiplets can be safely ignored and we will treat all components on an equal footing in our analysis.
The choice of $M_3 \gg M_2$ is for convenience, so that the decay and washout of $N_3$ may be treated separately from the $N_2$ stage (or indeed, to a very good approximation ignored), as well as ensuring that the resulting CP asymmetry due to $N_2$ decays is not suppressed by $(M_1/M_2)^2$ for certain choices of parameters. At this point, we shall not demand that $M_2\gg M_\chi$ (or $M_\chi\gg M_1$) in (\ref{eqn:full_mass_rel}) although this choice remains a valid (and often favored) possibility.

Suppose our $N_2$ is indeed providing the main ingredient for baryogenesis, then the general relation between the predicted baryon-to-photon ratio, $\eta_B$ and the final $B-L$ asymmetry, $\mathcal{N}_{B-L}^\text{f}$ is given by
\beqn{eqn:etaB_N_B-L}
 \eta_B \simeq 0.76\times 10^{-2} \mathcal{N}_{B-L}^\text{f}\;,
\eeqn
where we have used the notation $\mathcal{N}_X$ to denote the number of particle in quantity $X$ (in this case $X\equiv B-L$) per comoving volume which contains exactly one photon at some temperature much greater than the leptogenesis temperature.\footnote{This is an alternative way (c.f. quantity $Y_1$ in (\ref{eqn:BE_DM})) to keep track of the number density of a particle species during the evolution of the early universe.} The normalization for this volume is such that in thermal equilibrium ($T\gg M_2$), there is $\mathcal{N}_{N_2}^\text{eq} =1$ particle for each component of $N_2$.
Our prefactor of $0.76\times 10^{-2}$, which takes into account the dilution from sphaleron conversion and the expansion of the universe, differs (very slightly) from the $0.96\times 10^{-2}$ by other authors because we have an increased number of relativistic degrees of freedom coming from the multiplets $\chi$ and $N_{1,2}$. 

Successful leptogenesis simply means that the result in (\ref{eqn:etaB_N_B-L}) must match the corresponding measured valued from WMAP: $\eta_B^\text{CMB} =(6.19\pm 0.15) \times 10^{-10}$ 
\cite{Komatsu:2010fb}. 
In the ``$N_2$-leptogenesis'' scenario, this roughly translates into two basic requirements. Firstly, the CP asymmetry due to the out-of-equilibrium decays of $N_2$ (denoted $\varepsilon_2$) must be sufficient (and preferably with $\varepsilon_1$ from $N_1$ decays suppressed). Secondly, the $N_2$ generated asymmetry must (partially) evade the washout interactions involving the $N_1$'s after production. Usually, the first condition is easily satisfied with plenty of parameter space freedom while the second condition is more subtle and, as recently pointed out in \cite{Antusch:2010ms, Bertuzzo:2010et}, flavor effects in leptogenesis play a vital role.

\begin{figure}[tb]
\begin{center}
\textbf{(a)}
\includegraphics[width=0.23\columnwidth]{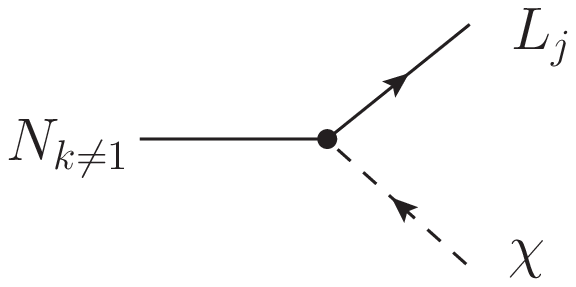}
\textbf{(b)}
\includegraphics[width=0.25\columnwidth]{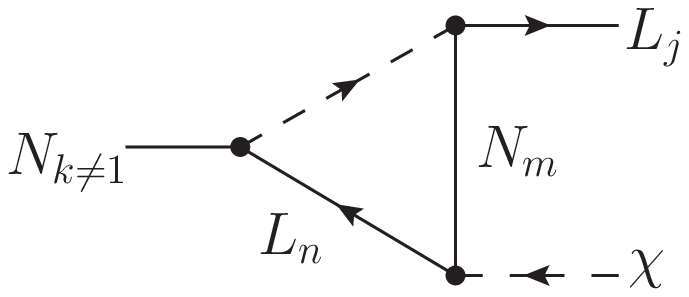}
\textbf{(c)}
\includegraphics[width=0.28\columnwidth]{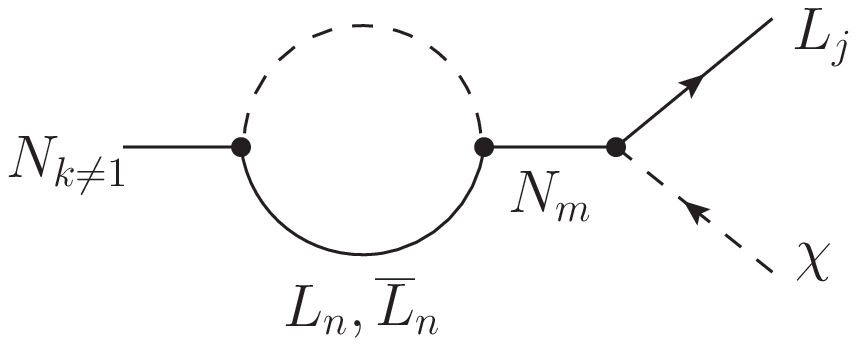}
\caption{The lepton number violating \textbf{(a)} tree-level, \textbf{(b)} one-loop vertex and \textbf{(c)} one-loop self-energy diagrams for $N_k$ multiplet decays ($k\neq 1$). For $k=1$, these processes are kinematically forbidden.}
\label{fig:tree_vsloop}
\end{center}
\end{figure}

To enunciate how our model fits inside the grand picture, let us recall the interaction Lagrangian in (\ref{eqn:main_Lag}) and expand out the relevant terms in component form:
\begin{align}
 -2\sqrt{5}\,\mathcal{L}_\text{int}' &=
  M_k \left[ 
    \overline{(N^{++}_k)^c} N_k^{--}
    -\overline{(N^{+}_k)^c} N_k^{-}
    +\overline{(N^{0}_k)^c} N_k^{0}
    -\overline{(N^{-}_k)^c} N_k^{+}
    +\overline{(N^{--}_k)^c} N_k^{++}
  \right] \nonumber 
  \\
  &\qquad
    +\sqrt{2}\, h_{jk}\,\overline{\nu_j^0} \left[
      \frac{1}{\sqrt{3}} \chi^{--} N_k^{++}
     -\sqrt{\frac{2}{3}} \chi^{-} N_k^{+}
     + \chi^0 N_k^0
     -\frac{2}{\sqrt{3}} \chi^{+} N_k^{-}
     +\sqrt{\frac{5}{3}} \chi^{++} N_k^{--}
    \right]\nonumber 
  \\
  &\qquad
    +\sqrt{2}\,  h_{jk}\,\overline{\ell_j^-} \left[
      \sqrt{\frac{5}{3}} \chi^{---} N_k^{++}
     -\frac{2}{\sqrt{3}}  \chi^{--} N_k^{+}
     + \chi^- N_k^0
     -\sqrt{\frac{2}{3}}\chi^{0} N_k^{-}
     +\frac{1}{\sqrt{3}} \chi^{+} N_k^{--}
    \right]
    +h.c.\;.
    \label{eqn:L_mass_expanded}
\end{align}
The second and third lines will induce the necessary lepton number violating processes that are needed for leptogenesis. Each component of $N_k$ (for $k\neq 1$) can decay into a lepton plus a corresponding component of $\chi$ (see Fig.~\ref{fig:tree_vsloop}a). After summing (incoherently) the contributions from each component, one obtains the following total rate at tree-level
\beqn{eqn:N_decay_rate}
 \Gamma(N_k \rightarrow  L_j\chi^\dagger)
 \equiv
 \Gamma(\overline{N}_k \rightarrow  \overline{L}_j \chi)
 = \frac{h_{jk}^* h_{jk}}{16\pi}\,M_k \left( 1-\frac{M_\chi^2}{M_k^2}
  \right)^2 ,\quad k \neq 1\;.
\eeqn
The CP asymmetry for the decay of $N_{k\neq 1}$ into $\chi$ and a lepton of flavor $j$ is defined as
\beqn{eqn:CP_defn}
 \varepsilon_{kj} \equiv 
  \frac{\Gamma(N_k\rightarrow L_j \chi^\dagger)
           -\Gamma(\overline{N}_k\rightarrow \overline{L}_j \chi)}
  {\sum_j \Gamma(N_k\rightarrow L_j \chi^\dagger)
           +\sum_j \Gamma(\overline{N}_k\rightarrow \overline{L}_j \chi)}
           \;.
\eeqn
The leading contribution to $\varepsilon_{kj}$ is obtained from the interference between the tree-level (Fig.~\ref{fig:tree_vsloop}a) and one-loop correction graphs  (Fig.~\ref{fig:tree_vsloop}b and Fig.~\ref{fig:tree_vsloop}c), and we get
\beqn{eqn:eps_cp_main}
 \varepsilon_{kj} =  \frac{1}{8\pi \left(h^\dagger h\right)_{kk}} 
 \sum_{m\neq k} \frac{ \text{Im}\left[h_{jk}^* h_{jm} (h^\dagger h)_{km}\right] f_1(M_k,M_m,M_\chi)
 +
 \text{Im}\left[h_{jk}^* h_{jm} (h^\dagger h)_{mk}\right] f_2(M_k,M_m,M_\chi)}
 {M_k^3 (M_k^2-M_m^2)(M_k^2-M_\chi^2)^2}\;,
\eeqn
where
\begin{align}
 f_1(M_k,M_m,M_\chi) &= M_m (M_k^2 - M_\chi^2)^2 
  \left[2 M_k^4 -M_k^2(M_m^2+2M_\chi^2) +M_\chi^4\right] \nonumber
  \\
  &\qquad
  +M_k^4 M_m (M_m^2 - M_k^2)(M_k^2+M_m^2 -2M_\chi^2) \,
  \ln \left[\frac{M_k^2(M_k^2+M_m^2 -2 M_\chi^2)}{M_k^2 M_m^2 - M_\chi^2} \right]\;,
  \\
 f_2(M_k,M_m,M_\chi) &= M_k (M_k^2 - M_\chi^2)^2 
  (M_k^4 - 2M_k^2 M_\chi^2 + M_\chi^4)\;.
\end{align}
Note that in the limit of $M_k\gg M_\chi$, both (\ref{eqn:N_decay_rate}) and (\ref{eqn:eps_cp_main}) reduce to the form that is identical to the results of standard leptogenesis \cite{Covi:1996wh}.

An important observation about this model is that although $N_1$ cannot decay via coupling $h_{j1} \overline{L}_j \,\chi\, N_1$, hence $\varepsilon_{1j}=0$ automatically, 6-plet $\chi$ may undergo the $L$-violating process $\chi \rightarrow L_j \overline{N}_1$ instead. The corresponding decay rate would be in the same form as (\ref{eqn:N_decay_rate}) but with the replacements: $M_k\rightarrow M_\chi$ and $M_\chi \rightarrow M_1$.
Therefore, the role of ``$N_1$ washout'' (in standard ``$N_2$-leptogenesis'') has essentially been taken up by field $\chi$ in our scenario. 

However, there can be no CP asymmetry generated from the decay of 6-plet $\chi$. It is because there is only \emph{one} copy of $\chi$ in the model, and thus the internal $\chi$ running in a loop diagram, similar to those depicted in Fig.~\ref{fig:tree_vsloop}b and Fig.~\ref{fig:tree_vsloop}c, must be the same as the external one. Consequently, the interference term will have a vanishing absorptive part.
This means, the total $B-L$ asymmetry must originate from our 5-plet $N_2$ decays entirely,\footnote{We will neglect the suppressed contribution from $\varepsilon_{3j}$.} making this a classic ``$N_2$-leptogenesis''-like scenario.
 
To make our following illustration more definite, let us specialize to a hierarchical mass spectrum for our exotic particles, namely, 
$M_3\gg M_2 \gg M_\chi \gg M_1 \simeq 10$~TeV. In addition to this, we will assume 
\beqn{eqn:mass_scale_set}
M_3 \gg 10^{12}\text{ GeV}\;,\quad 
10^{12}\text{ GeV} \gg M_2 \gg 10^{9}\text{ GeV}\;,\quad  
10^{9}\text{ GeV} \gg M_\chi \gg M_1 \;.
\eeqn
These restrictions are chosen so that flavor effects \cite{Vives:2005ra,flavor_eff, flavor_eff_Nardi, JosseMichaux:2007zj, Antusch:2010ms, Bertuzzo:2010et} in leptogenesis can be treated in a more straight forward manner and it is in no way a model building requirement. 

After disregarding the unimportant effects from $M_3$ decays at very high temperatures, there are basically \emph{two} major stages of interest in this kind of setup. The first is the asymmetry production stage at $T\sim M_2$ where $N_2$ decays and related washout processes are active while the second is the washout stage at $T\sim M_\chi$ where effects from $M_\chi$ inverse decays become prominent. We will discuss them in turn.

\subsection{Asymmetry production stage at $T\sim M_2$}

In general, at $T \ll 10^{12}$~GeV, flavor effects become important as the charged lepton Yukawa interactions in equilibrium essentially introduce a source of decoherence whereby the lepton state $|L_{(k)}\rangle$, generated by the decays of $N_k$'s,\footnote{The use of the subscript $(k)$ is to remind us that the flavor decomposition of $L_{(k)}$ can be different for each $k$.} are projected onto one of the three flavor eigenstates, $|L_j\rangle, j=e,\mu,\tau$ with probability $|\langle L_{(k)}|L_j\rangle|^2$. Therefore, at our asymmetry production stage, $T\sim M_2$, this effect must be properly accounted for.

However, since $M_2\gg 10^9$~GeV, only the tauon Yukawa interaction (from $\overline{L}_\tau\,\phi\,e_{R\tau}$) is in equilibrium. As a result, we effectively has a two-flavor situation where $|L_{(2)}\rangle$ will either be projected onto $|L_{\tau}\rangle$ or its orthogonal state $|L_{\perp}\rangle$ (which is a coherent linear combination of the $e$ and $\mu$ flavor states).\footnote{Note that the mass ranges in (\ref{eqn:mass_scale_set}) was carefully selected such that these lepton decoherence interactions are either fully in-equilibrium or out-of-equilibrium but not in between. This makes the handling of flavor effects a lot more transparent.} To track the evolution of the lepton asymmetry for each flavor state, it is convenient to introduce the flavor projectors
\beqn{eqn:flavor_proj}
 P_{2j} \equiv \frac{\Gamma_{2j}}{\Gamma_2} = |\langle L_{(2)}|L_j \rangle|^2 \;,
 \quad
  \overline{P}_{2j} \equiv \frac{\overline{\Gamma}_{2j}}{\overline{\Gamma}_2} 
  = |\langle \overline{L}_{(2)}|\overline{L}_j \rangle|^2\;,
\eeqn
where $\Gamma_{2j} = \Gamma(N_2\rightarrow L_j\chi^\dagger)$,
$\overline{\Gamma}_{2j} = \Gamma(\overline{N}_2\rightarrow \overline{L}_j\chi)$ and $\Gamma_2, \overline{\Gamma}_2$ are the corresponding rates with $j$ summed over. Note that states $|L_{(2)}\rangle$ and $|\overline{L}_{(2)}\rangle$ are \emph{not} CP conjugates of each other in general. In the current case, flavor $j = \tau$ or $\perp$. The tree-level contribution to the quantities in (\ref{eqn:flavor_proj}) is given by
\beqn{eqn:flavor_proj_0}
 P_{2j}^0 \equiv \overline{P}_{2j}^0 = 
 \frac{h_{j2}^* h_{j2}}{(h^\dagger h)_{22}}\;,
\eeqn
where $h$ is the Yukawa coupling in (\ref{eqn:main_Lag}). With this, we may write down the set of evolution equations for $N_2$ and lepton asymmetry $\Delta_j$ in flavor $j$ as
\begin{align}
 \frac{d\mathcal{N}_{N_2}}{dz} &=
   -D_2 \left( \mathcal{N}_{N_2} - \mathcal{N}_{N_2}^\text{eq}
   \right)\;,
\label{eqn:BE_z2_N2}
\\
 \frac{d\mathcal{N}_{\Delta_\perp}}{dz} &=
  -\varepsilon_{2\perp} D_2 \left( \mathcal{N}_{N_2} -
  \mathcal{N}_{N_2}^\text{eq}   \right)
   - P^0_{2\perp} W_2 \sum_{j=\perp,\tau} C^{f=2}_{\perp j} \mathcal{N}_{\Delta_\perp} \;,
\label{eqn:BE_z2_Ltaup}
\\
 \frac{d\mathcal{N}_{\Delta_\tau}}{dz} &=
 -\varepsilon_{2\tau} D_2 \left( \mathcal{N}_{N_2} -
  \mathcal{N}_{N_2}^\text{eq}   \right)
   - P^0_{2\tau} W_2 \sum_{j=\perp,\tau} C^{f=2}_{\tau j} \mathcal{N}_{\Delta_\tau} \;,
\label{eqn:BE_z2_Ltau}
\end{align}
where $z= M_2/T$. The decay ($D_2$) and washout ($W_2$) terms are given by
\beqn{eqn:decay_washout_defn}
 D_2 = \frac{\left(\Gamma_{2}+\overline{\Gamma}_{2}\right)}{z\, H(z)}
       \frac{\mathcal{K}_1(z)}{\mathcal{K}_2(z)}\;,
       \qquad
 W_2 = \frac{1}{2}\, D_2 \, \frac{\mathcal{N}_{N_2}^\text{eq}}
    {\mathcal{N}_{L}^\text{eq} \mathcal{N}_{\chi}^\text{eq}}      \;,
\eeqn
with $\mathcal{K}_n(z)$ denoting the $n$th order modified Bessel function of the second kind and $H(z)$ being the Hubble expansion rate at $z$. For our particular normalization, $\mathcal{N}_{N_2}^\text{eq} = z^2 \mathcal{K}_2(z)/2$, while $\mathcal{N}_{L}^\text{eq} =1$. Note that as in the typical analysis for leptogenesis, only the dominant contribution to washout $W_2$, namely the $N_2$ inverse decay process, has been included in (\ref{eqn:decay_washout_defn}).\footnote{It is worth pointing out that in our model, scattering processes that are mediated by 6-plet $\chi$ do not exist as $\chi$ does not couple to quarks, whereas you can still have interactions like $\chi\chi \leftrightarrow L_i L_j$ mediated by 5-plet $N_k$ that can contribute to washout. But as usual, their effects are subdominant.}

The flavor coupling matrix $C^{f=2}$ which links $\mathcal{N}_{\Delta_\perp}$ and $\mathcal{N}_{\Delta_\tau}$ is given by
\cite{JosseMichaux:2007zj,flavor_eff_Nardi,Antusch:2010ms}
\beqn{eqn:flavor_C2}
 C^{f=2} \equiv 
 \twtwMat{C^{f=2}_{\perp\perp}}{C^{f=2}_{\perp\tau}}
          {C^{f=2}_{\tau\perp}}{C^{f=2}_{\tau\tau}} 
 =\twtwMat{417/589}{-120/589}
          {-30/589}{390/589}\;.
\eeqn
Unlike \cite{Antusch:2010ms}, our flavor coupling matrix contains only the part related to the asymmetry stored in the lepton sector (like the one used in \cite{JosseMichaux:2007zj}), and not the part involving the effects from Higgs $\phi$. This is simply because scalar $\phi$ does not participate in the $N_2 \rightarrow L_j \chi^\dagger$ process. But, as noted in \cite{Antusch:2010ms}, neglecting the $\phi$ part will not change the qualitative result since what is essential here is that the off-diagonal terms in  (\ref{eqn:flavor_C2}) are nonzero. Consequently, enhancement to the final lepton asymmetry after this stage (which is simply given by $\mathcal{N}_{B-L}^{T_z}=\mathcal{N}_{\Delta_\perp}^{T_z}+\mathcal{N}_{\Delta_\tau}^{T_z}$, where $T_z$ denotes the temperature at the end of this period)
becomes possible just like in regular ``$N_2$-leptogenesis'' with flavor effects turned on. 

Explicitly, the quantities $\mathcal{N}_{\Delta_\perp}^{T_z}$ and $\mathcal{N}_{\Delta_\tau}^{T_z}$ can be calculated by first applying the transformation
\beqn{eqn:prod_U_trans}
\twoneMat{\widetilde{\mathcal{N}}_{\Delta_\perp}}
 {\widetilde{\mathcal{N}}_{\Delta_\tau}}
 =U_2 \twoneMat{\mathcal{N}_{\Delta_\perp}}{\mathcal{N}_{\Delta_\tau}}
 \;,\qquad
 \twoneMat{\widetilde{\varepsilon}_{2\perp}}
 {\widetilde{\varepsilon}_{2\tau}}
 =U_2 \twoneMat{\varepsilon_{2\perp}}
 {\varepsilon_{2\tau}}
 \;,
\eeqn
to make them decoupled in Eqs. (\ref{eqn:BE_z2_Ltaup}) and (\ref{eqn:BE_z2_Ltau}). In other words, $U_2$ diagonalizes
\beqn{eqn:U_2_diag}
 U_2 \twtwMat{P^0_{2\perp}C^{f=2}_{\perp\perp}}
                            {P^0_{2\perp}C^{f=2}_{\perp\tau}}
          {P^0_{2\tau}C^{f=2}_{\tau\perp}}
                               {P^0_{2\tau}C^{f=2}_{\tau\tau}}
   U_2^{-1}
   =
   \twtwMat{\widetilde{P}^0_{2\perp}}{0}{0}{\widetilde{P}^0_{2\tau}}\;,
\eeqn
and we obtain
\beqn{eqn:BE_z2_decoupled}
  \frac{d\widetilde{\mathcal{N}}_{\Delta_j}}{dz} =
  -\widetilde{\varepsilon}_{2j}\, D_2 \left( \mathcal{N}_{N_2} -
  \mathcal{N}_{N_2}^\text{eq}   \right)
   - \widetilde{P}^0_{2j} \,W_2\, \widetilde{\mathcal{N}}_{\Delta_j} \;,\qquad
   j=\perp, \tau\;.
\eeqn
Subsequently, (\ref{eqn:BE_z2_decoupled}) together with (\ref{eqn:BE_z2_N2}) can be solved as per usual \cite{listwashout, Vives:2005ra, flavor_eff_Nardi, flavor_eff, JosseMichaux:2007zj, Antusch:2010ms} and one eventually gets\footnote{We do not dwell on the subtleties related to initial conditions here as it is a grander issue in standard leptogenesis with flavor effects. See for example \cite{Bertuzzo:2010et}.}
\beqn{eqn:N_B-L_Tz}
 \widetilde{\mathcal{N}}_{B-L}^{T_z}=  
 \sum_{j=\perp,\tau}
    \widetilde{\mathcal{N}}_{\Delta_j}^{T_z}
\simeq \sum_{j=\perp,\tau}
  \widetilde{\varepsilon}_{2j} \; \kappa(\widetilde{P}^0_{2j} K_{z})\;,
\eeqn
where the final efficiency factor (for an initial thermal abundance of $N_2$) has the form
\beqn{eqn:eff_factor_2}
 \kappa(X) \simeq \frac{2}{z_B X}
 \left(1- e^{-z_B X/2}\right)\;, \quad 
 z_B \simeq 2+ 4X^{0.13} \,e^{-2.5/X}
 \simeq 
 \mathcal{O}(1) \text{ to }\mathcal{O}(10)\;. 
\eeqn
The total decay parameter, $K_z$, for the $N_2$ process is defined as
\beqn{eqn:total_decay_Kz}
 K_z \equiv \frac{\left(\Gamma_{2}+\overline{\Gamma}_{2}\right)}{H(z=1)}\;,
\eeqn
and the corresponding flavored version is given by $K_{zj} \equiv P^0_{2j} K_z$ such that $\sum_j K_{zj} = K_z$.

Since result (\ref{eqn:N_B-L_Tz}) is written in terms of the $U_2$-rotated quantities, one must carefully reverse the procedure to obtain the desired $\mathcal{N}_{B-L}^{T_z}$. However, we will not show these steps explicitly. The main point is to see how the parameters of our multiplet model enter into the theory of ``$N_2$-leptogenesis'' for the production stage.

\subsection{Washout stage at $T\sim M_\chi$}

As aforementioned, no new lepton asymmetry will be produced at this stage, and thus the total asymmetry can only decrease from here on. However, owing to flavor effects, such suppression may not always be effective. This is because the asymmetry stored in each flavor are washed out differently when flavor effects are considered, and as a consequence, some part of the asymmetry may evade the washout entirely, making it possible to have successful ``$N_2$-leptogenesis''.

Since we have assumed that $M_\chi \ll 10^9$~GeV, it is inevitable that the lepton asymmetry, $\mathcal{N}_{\Delta_\perp}^{T_z}$, will eventually decohere into the $\mu$- and $e$-flavor components as muon Yukawa interactions come into equilibrium at $T\lesssim 10^9$~GeV. With the tauon Yukawa processes already in equilibrium at this temperature, we therefore have the full three-flavor case \cite{Vives:2005ra,flavor_eff, flavor_eff_Nardi}.

Suppose $\mathcal{N}_{\Delta_\perp}^{T_z}$ breaks into the $\mu$- and $e$-components \emph{before} the $M_\chi$ related washout effects become prominent, so the situation is more clear-cut. In other words, let us assume there exists a temperature $T'$ such that $M_\chi \ll T'\ll 10^9$~GeV, and for which the decoherent effect mentioned can begin to happen. It is not hard to see that there is a plenty of parameter space freedom in our model to allow for this.
With this assumption, one can estimate the individual lepton asymmetry for the  $\mu$- and $e$-flavor at $T'$ \cite{Antusch:2010ms}:\footnote{Note that $\mathcal{N}_{\Delta_\tau}^{T'} \simeq \mathcal{N}_{\Delta_\tau}^{T_z}$.}
\beqn{eqn:emu_asy_at_Tprime}
 \mathcal{N}_{\Delta_i}^{T'} \simeq \left(\varepsilon_{2i} 
 -\frac{P^0_{2i}}{P^0_{2\perp}}\,\varepsilon_{2\perp}
 \right)\mathcal{N}_{N_2}^{T\gg T_z}
 +\frac{P^0_{2i}}{P^0_{2\perp}}\,\mathcal{N}_{\Delta_\perp}^{T_z}\;,
 \qquad i = e, \mu\;,
\eeqn
where the expression in the first term is highly sensitive to initial conditions.

Subsequently, when the $\chi \;(\rightarrow L\overline{N}_1)$ inverse decay dominated washout\footnote{Scattering interactions mediated by $N_1$ (e.g. $\chi\chi \leftrightarrow L_i L_j$) have been ignored.} kicks in at $T\sim M_\chi$, we have the following set of evolution equations for the asymmetry in flavor $j$:
\beqn{eqn:BE_chi_washout}
  \frac{d\mathcal{N}_{\Delta_j}}{dx} =
   - P^0_{\chi j}\, W_\chi \sum_{i=e,\mu,\tau} C^{f=3}_{ji} \mathcal{N}_{\Delta_j} \;, \qquad j = e,\mu,\tau\;,
\eeqn
where $x=M_\chi/T$ and 
\beqn{eqn:projX0__WX}
 P_{\chi j}^0 \equiv \overline{P}_{\chi j}^0 = 
 \frac{h_{j1}^* h_{j1}}{(h^\dagger h)_{11}}\;, \qquad
 W_\chi = \frac{\Gamma_\chi^\text{ID}}{x\, H(x)}
           \frac{\mathcal{K}_1(x)}{\mathcal{K}_2(x)}
           \frac{\mathcal{N}_{\chi}^\text{eq}}
           {\mathcal{N}_L^\text{eq}\mathcal{N}_{N_1}^\text{eq}}\;,
\eeqn
with $\Gamma_\chi^\text{ID}$ denoting the tree-level inverse decay rate ($L \overline{N}_1 \rightarrow \chi$), which is given by $(h^\dagger h)_{11} M_\chi /(16\pi)$ when $M_\chi \gg M_1$. The 3-flavor coupling matrix is \cite{JosseMichaux:2007zj,flavor_eff_Nardi,Antusch:2010ms}
\beqn{eqn:flavor_C3}
 C^{f=3} \equiv\ththMat{C^{f=3}_{ee}}{C^{f=3}_{e\mu}}{C^{f=3}_{e\tau}}
                     {C^{f=3}_{\mu e}}{C^{f=3}_{\mu\mu}}{C^{f=3}_{\mu\tau}}
                   {C^{f=3}_{\tau e}}{C^{f=3}_{\tau\mu}}{C^{f=3}_{\tau\tau}}
    =\ththMat{151/179}{-20/179}{-20/179}
             {-25/358}{344/537}{-14/537}
             {-25/358}{-14/537}{344/537}
             \;,
\eeqn
where again it contains only the lepton part but not the $\phi$ part. Analogous to the previous section, one may define the transformation $U_3$ so that
\beqn{eqn:U_3_diag}
 U_3 \ththMat{P^0_{\chi e}C^{f=3}_{ee}}
                     {P^0_{\chi e}C^{f=3}_{e\mu}}
                            {P^0_{\chi e}C^{f=3}_{e\tau}}   
          {P^0_{\chi\mu}C^{f=3}_{\mu e}}
                     {P^0_{\chi\mu}C^{f=3}_{\mu\mu}}
                            {P^0_{\chi\mu}C^{f=3}_{\mu\tau}}
          {P^0_{\chi\tau}C^{f=3}_{\tau e}}
                     {P^0_{\chi\tau}C^{f=3}_{\tau\mu}}
                            {P^0_{\chi\tau}C^{f=3}_{\tau\tau}}
   U_3^{-1}
   =
   \ththMat{\widehat{P}^0_{\chi e}}{0}{0}
   {0}{\widehat{P}^0_{\chi\mu}}{0}
   {0}{0}{\widehat{P}^0_{\chi\tau}}
   \;.
\eeqn
Then, the general solution for (\ref{eqn:BE_chi_washout}) in the $U_3$-rotated variables (denoted with a ``hat'') is 
\beqn{eqn:BE_washout_soln}
 \widehat{\mathcal{N}}^\text{f}_{\Delta_j}   
     \simeq \widehat{\mathcal{N}}^{T'}_{\Delta_j} \,e^{-\frac{3}{8\pi} 
     \widehat{P}^0_{\chi j} K_x}
     \;,\qquad 
 j= e,\mu,\tau \;,
\eeqn
where the total decay parameter, $K_x \equiv 2\Gamma_\chi/ H(x=1)$, and the $j$-flavored case is defined as $K_{xj} \equiv P^0_{\chi j} K_x$. The total $B-L$ asymmetry can be obtained by summing the contributions from all flavors after carefully transforming the $U_3$-rotated variables back into the proper basis.\\ 

At this point, it is obvious that our multiplet model is effectively a carbon copy of the standard flavored ``$N_2$-leptogenesis'' scenario as far as solving the baryogenesis problem is concerned. However, the only constraint we have used so far is the lower mass bound from DM considerations. It remains to be confirmed if the elements of the Yukawa matrix $h$ can be chosen such that they are also compatible with neutrino masses and mixings. In the next section, we will focus on the constraints coming from neutrino phenomenologies.

\section{Neutrino masses from radiative seesaw}\label{sec:Rseesaw}

\begin{figure}[tb]
\begin{center}
\includegraphics[width=0.38\columnwidth]{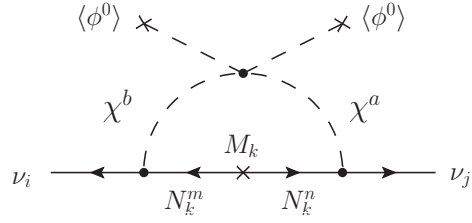}
\caption{The one-loop diagram for generating light neutrino mass term $(M_\nu)_{ij}$. $(N^m_k,N^n_k)$ and $(\chi^a,\chi^b)$ are pairs of components from multiplets $N_k$ and $\chi$ respectively. They are chosen such that charge conservation is obeyed at all vertices. As a result, the combined electric charge within each pair must be zero, i.e. $Q(N^m_k) = -Q(N^n_k)$ and $Q(\chi^a)=-Q(\chi^b)$. For example, one possible set of choices is 
$(N_k^m,N_k^n) = (N_k^+,N_k^-)$ and $(\chi^a,\chi^b) = (\chi^+,\chi^-)$.}
\label{fig:nuMass}
\end{center}
\end{figure}

Recall from the discussion in Sec.~\ref{sec:model} that in order to have a DM candidate in this model, 6-plet $\chi$ must not develop a nonzero VEV. Consequently, Yukawa coupling $\overline{L}_j\,\chi\,N_k$ cannot give rise to a Dirac mass term after spontaneous symmetry breaking, and light neutrinos will remain massless at tree-level. However, as hinted earlier, neutrino mass terms can be generated radiatively in this model via the one-loop diagrams as depicted in Fig.~\ref{fig:nuMass}.

The three relevant interaction terms from Lagrangian (\ref{eqn:main_Lag}) that enter into Fig.~\ref{fig:nuMass} are the Yukawa, $h_{jk}\overline{L}_j\chi N_k$, the Majorana term, $\overline{(N_k)^c} M_k N_k$, and the quartic coupling, $\lambda_{\phi\chi}'(\phi\chi)^2$ (plus their hermitian conjugate).
Since $N_k$ and $\chi$ are multiplets, there are several combinations of component fields which are allowed in the loop for each $k$. For instance in Fig.~\ref{fig:nuMass}, the pair ($N_k^m , N_k^n$) could be ($N_k^{++},N_k^{--}$),  ($N_k^{+},N_k^{-}$) or ($N_k^{0},N_k^{0}$), and the cases with the order for $m$ and $n$ swapped. But because of the requirement of electric charge conservation, once the pair ($N_k^m , N_k^n$) is chosen, there is only one combination of ($\chi^a ,\chi^b$) that is compatible (see Fig.~\ref{fig:nuMass}).

To calculate all these contributions, one must again work in the component form for (\ref{eqn:main_Lag}). The expansion for the relevant Yukawa and Majorana terms was presented in the first two lines of (\ref{eqn:L_mass_expanded}) while for the quartic term $\lambda_{\phi\chi}'(\phi\chi)^2$, it is (after spontaneous symmetry breaking)
\beqn{eqn:lambda_HHXX_Lag}
\mathcal{L}_{\phi\phi\chi\chi}=
 \frac{\sqrt{5}}{12}\,\lambda_{\phi\chi}'\left[ \frac{2}{\sqrt{5}}\,\chi^{++}\chi^{--}
  -\frac{4\sqrt{2}}{5} \chi^+\chi^-
  +\frac{3}{5} \chi^0\chi^0
 \right]  \langle\phi^0\rangle^2 
 +h.c.\;.
\eeqn
Neglecting the tiny mass splittings between the components in each multiplet 
(hence letting  $M_{N_k^m} = M_{N_k^n} = M_k$ and $M_{\chi^a}=M_{\chi^b} = M_\chi$), and after summing over all contributions, the light neutrino mass matrix is found to be
\beqn{eqn:nuMass_exp}
 (M_\nu)_{ij} = \frac{7}{5760\pi^2}\,
 \lambda_{\phi\chi}'\langle\phi^0\rangle^2
 \sum_k\frac{h_{ik} h_{jk}}{M_k}
 \,
 \frac{y_k}{1-y_k}\left[1+\frac{y_k \ln y_k}{1-y_k}
 \right],
 \qquad \text{where }\;\; y_k\equiv \frac{M_k^2}{M_\chi^2}\;.
\eeqn
Specializing in the mass spectrum of (\ref{eqn:mass_scale_set}) where $M_1\ll M_\chi\ll M_{2,3}$, we have $y_1\ll 1$ and $y_{2,3}\gg 1$. So, (\ref{eqn:nuMass_exp}) may be approximated as
\beqn{eqn:nuMass_exp_approx}
 (M_\nu)_{ij} \simeq \frac{7}{5760\pi^2}\,
 \lambda_{\phi\chi}'\langle\phi^0\rangle^2
 \left[
  \frac{h_{i1} h_{j1}}{M_1}\,y_1
  +\sum_{k=2,3} \frac{h_{ik} h_{jk}}{M_k}\,\left(\ln y_k -1 \right)
 \right]\;.
\eeqn
The individual light neutrino mass is obtained by diagonalizing $M_\nu$ with the neutrino mixing matrix, $U_\text{PMNS}$:
\beqn{eqn:nuMass_diag}
 m^\nu_{n} = \sum_{i,j}\, (U_\text{PMNS}^\dagger)_{ni} \,
  (M_\nu)_{ij} \,
  (U_\text{PMNS}^* )_{jn}
  \;,
 \qquad n=1,2,3\;.
\eeqn
An important observation here is that the light neutrino mass scale will be dependent on the size of $\lambda_{\phi\chi}'$, as well as the seesaw factors of $h_{ik} h_{jk} \langle\phi^0\rangle^2 /M_k$. Moreover, recall that $\lambda_{\phi\chi}'$ is not constrained by the vacuum stability issue discussed in Sec.~\ref{sec:model} and thus it is essentially a free parameter. This fact will allow a very interesting situation where for a given $M_k$ scale, the size of the Yukawas $h$'s can be large so that observable lepton flavor violating effects in processes such as $\mu\rightarrow e \gamma$ and $\mu$-$e$ conversion in atomic nuclei are possible without spoiling the light neutrino mass predictions. A recent analysis on this point can be found in \cite{XGMDM}.

\section{Discussion on the parameter space}

The stage is now set for us to combine the results of the last three sections and explicitly show that there exists a parameter space for our exotic multiplet model to simultaneously solve the problems of dark matter, baryon asymmetry and neutrino masses.

Firstly, as far as the dark matter condition goes, the key constraint is coming from Eq.~(\ref{eqn:M_1_mass}) which sets the lower bound for the dark matter candidate, $N_1$ at $M_1\simeq 10$~TeV.\footnote{Strictly speaking, the vacuum stability conditions of the scalar potential (as discussed in Sec.~\ref{sec:model}) provides further constraints on the DM parameter space (albeit for $\lambda_{\chi\alpha}$ and $\lambda_{\phi\chi\beta}$ only  which do not directly affect other sectors of the theory). However, since these are higher-order effects, we do not incorporate them in our analysis explicitly, and shall assume that they can be properly fine-tuned if required.}
Note that all our subsequent discussions in the previous two sections were based on this assumption, therefore, our model will automatically satisfy the DM condition regardless of how we may choose the parameters in other sectors of the theory.

Next, observe that (\ref{eqn:N_decay_rate}), (\ref{eqn:eps_cp_main}),
(\ref{eqn:flavor_proj_0}), (\ref{eqn:projX0__WX}) and the definition for $K_{z,x}$ are the most important quantities which govern the leptogenesis sector, while (\ref{eqn:nuMass_exp}) is the main result for neutrinos. The common features of all of them is that they are highly dependent on the Yukawa matrix elements, $h_{jk}$ and masses, $M_{k,\chi}$. For the neutrino sector, $\lambda_{\phi\chi}'$ is also essential. Thus, the task is basically to select a set of $h_{jk}, M_{k,\chi}$ and $\lambda_{\phi\chi}'$ that can produce successful baryogenesis while giving realistic neutrino masses and mixing. 

To see that there is a compatible set of choices, we begin by following condition (\ref{eqn:mass_scale_set}) and take 
\beqn{eqn:Mass_Choices}
 M_1 \simeq 10^{4} \text{ GeV}\;,\quad
  M_\chi \simeq 10^{7} \text{ GeV}\;,\quad
   M_2 \simeq 10^{10} \text{ GeV}\;,\quad
    M_3 \simeq 10^{13} \text{ GeV}\;.
\eeqn
For the light neutrino sector, we pick the mass scale to be at $m_0 \approx 0.002$~eV (normal or inverted hierarchy) and $m_0^\text{QD} \approx 0.16$~eV (quasi-degenerate). Appealing to the best fit values from neutrino oscillation experiments \cite{Schwetz:2008er}, 
\begin{eqnarray}\label{eqn:nu_data_mass_sq}
 &\Delta m_{12}^2 \approx 7.59 \times 10^{-5} \text{ eV}^2\;,\quad
 |\Delta m_{23}^2| \approx 2.40 \times 10^{-3} \text{ eV}^2\;,\\
 &\sin^2 \theta_{12} \approx 0.318\;,\quad
 \sin^2 \theta_{23} \approx 0.50\;,\quad
 \sin^2 \theta_{13} \approx 0.013\;,\label{eqn:nu_data_mixing}
\end{eqnarray}
one then has the following light neutrino mass spectra (in eV):
\begin{align}
 \label{eqn:NH_spectrum}
 \text{normal hierarchy: }\;\;
  &m_1 \approx 0.002\;, \quad  m_2 \approx 0.0089\;, \quad
    m_3 \approx 0.0498\;,\\
 \label{eqn:IH_spectrum}
 \text{inverted hierarchy: }\;\;
   &m_1 \approx 0.0482\;, \quad  m_2 \approx 0.0490\;, \quad
    m_3 \approx 0.002\;,\\
 \label{eqn:QD_spectrum}
 \text{quasi-degenerate: }\;\;
   &m_1 \approx 0.1600\;, \quad  m_2 \approx 0.1602\;, \quad
    m_3 \approx 0.1676\;.
\end{align}
We shall use these numbers for our analysis below.

\subsection{Normal hierarchy case}

To begin with, we take the spectrum of (\ref{eqn:NH_spectrum}) and reverse the diagonalization procedure in (\ref{eqn:nuMass_diag}) using  (\ref{eqn:nu_data_mixing}) as the inputs for $U_\text{PMNS}$ (ignoring the phases). So, we have for $\lambda_{\phi\chi}' = 0.1$
\beqn{eqn:M_nu_real_NH}
 \frac{5760\pi^2}{7\lambda_{\phi\chi}' \langle\phi^0\rangle^2}\, 
  M_\nu^\text{NH}
 \simeq
 \ththMat{1.287}{1.588}{-0.3707}
         {1.588}{7.404}{-5.697}
         {-0.3707}{-5.697}{7.601} \times 10^{-20} \text{ eV}^{-1}\;,
\eeqn
where $\langle\phi^0\rangle = 174$~GeV. To reproduce the entries in matrix (\ref{eqn:M_nu_real_NH}), one must find a set of $h_{jk}$'s such that Eq.~(\ref{eqn:nuMass_exp_approx}) is satisfied for the given exotic masses in (\ref{eqn:Mass_Choices}). But it is not hard to see that there are actually more parameters than constraints coming from matching (\ref{eqn:nuMass_exp_approx}) with (\ref{eqn:M_nu_real_NH}). This extra parameter space freedom is crucial because the conditions for successful ``$N_2$-leptogenesis'' with flavor effects necessarily restrict the pattern of $h_{jk}$ that one can choose.

Recall from earlier that flavor effect is the vital ingredient which opens up new parameter space for ``$N_2$-leptogenesis''. It hinges on the fact that asymmetry produced in a certain flavor may evade (sometimes completely) the washout at the ``$N_1$ stage'' (in our case, $N_1$'s role is played by 6-plet $\chi$). Therefore, to ensure successful leptogenesis in our particular setup, we must control the size of the flavor projectors, $P^0_{2j}$ and $P^0_{\chi j}$ in (\ref{eqn:flavor_proj_0}) and (\ref{eqn:projX0__WX}) respectively. This is usually done through the language of the flavored decay parameters $K_{zj}$ and $K_{xj}$, which have an one-to-one correspondence with (\ref{eqn:flavor_proj_0}) and (\ref{eqn:projX0__WX}).

Using the insights gained from some recent analyses of standard flavored leptogenesis \cite{JosseMichaux:2007zj,Antusch:2010ms,Bertuzzo:2010et}, we will demand that the final lepton asymmetry is mainly due to the effects in the $\tau$ flavor. In other words, we require $K_{z\tau} \gtrsim 1$ to be relatively big so that during the production stage at $T\sim M_2$, asymmetry $\Delta_{\tau}$ can be effectively produced,\footnote{A large $K_{z\tau}$ also means a stronger $\tau$-washout at the production stage. But at the same time, a large $K_{z\tau}$ (meaning large $K_z$) allows more $N_2$ to be produced thermally. So, the interplay between decay and washout is rather convoluted and one must explicitly solve the evolution equations in order to get a real feeling of what is large or small enough.} whereas at the $\chi$ washout stage, we need $K_{x\tau} \ll 1$ to be small so that the $\Delta_\tau$ produced from the previous stage can largely avoid washout and survive to make up the observed baryon asymmetry.

With this in mind, we solve our set of constraint equations from (\ref{eqn:M_nu_real_NH}) after setting 
\beqn{eqn:NH_Ks}
 K_{z\tau} \simeq 65 \;, \qquad
 K_{x\tau} \simeq 0.1 \;.
\eeqn
Note that these fix the magnitudes for $h_{2\tau}$ and $h_{1\tau}$ immediately after substituting in the value for $H(z=1)$ and $H(x=1)$ respectively. The Hubble parameters themselves are determined by 
\beqn{eqn:H_defn}
 H(w) \simeq 1.66 \sqrt{g_s^*}\,\frac{M_{w}^2}{M_\text{pl}}\frac{1}{w^2}\;, 
\eeqn
where $M_w = M_2$ and $g_s^* \approx 127.5$ for the $H(z=1)$ case, while  $M_w = M_\chi$ and $g_s^* \approx 115.5$ for the $H(x=1)$ case.  $M_\text{pl} \approx 1.22\times 10^{19}$~GeV is the Planck mass. The rest of the $h_{jk}$ entries may now be solved simultaneously, and one possible set of solutions for the complex Yukawas is
\begin{eqnarray}
 h_{e1} = 1.23+0.359i\;,\quad 
 h_{e2} = 0.104-0.329i\;,\quad 
 h_{e3} = -0.344+ 0.263i\;, \\
  h_{\mu 1} = 1.71-1.02 i\;,\quad 
  h_{\mu 2} = -0.304-0.468i\;,\quad 
  h_{\mu 3} = -3.76 + 0.367i\;,\\
   h_{\tau 1} = 1.07\times 10^{-5}\;,\quad 
   h_{\tau 2} = 8.88\times 10^{-3}\;,\quad 
   h_{\tau 3} = 5.34\;.\qquad\qquad
\end{eqnarray}
Using these in the calculations as outlined in Sec.~\ref{sec:lepto}, one will eventually find that $\eta_B \simeq 6.3\times 10^{-10}$ which agrees with the CMB data. It is interesting to point out that the inclusion of the nondiagonal flavor couplings $C^{f=2}$ and $C^{f=3}$ has increased the final asymmetry by about a factor of 2 in this particular example.

\subsection{Inverted hierarchy case}

One may repeat a similar analysis for the inverted hierarchical light neutrino spectrum of (\ref{eqn:IH_spectrum}). With the benefit of hindsight, we must choose a larger $\lambda_{\phi\chi}' = 1$ than before so that the resultant Yukawa couplings, $h_{jk}$ will at most be of \order{1}. The light neutrino mass matrix in this case is  
\beqn{eqn:M_nu_real_IH}
 \frac{5760\pi^2}{7\lambda_{\phi\chi}' \langle\phi^0\rangle^2}\, 
  M_\nu^\text{IH}
 \simeq
 \ththMat{12.8}{-0.931}{1.07}
         {-0.931}{6.88}{6.19}
         {1.07}{6.19}{6.90} \times 10^{-21} \text{ eV}^{-1}\;.
\eeqn
Like before, we will restrict the $\tau$-flavor decay parameters to obey $K_{z\tau}\gtrsim 1$ and $K_{x\tau} \ll 1$. One suitable choice for this is
\beqn{eqn:IH_Ks}
 K_{z\tau} \simeq 2 \;, \qquad
 K_{x\tau} \simeq 0.01 \;.
\eeqn
Subsequently, the Yukawas may be chosen as
\begin{eqnarray}
 h_{e1} = 2.73-2.63i\;,\quad 
 h_{e2} = -0.737-0.758i\;,\quad 
 h_{e3} = 0.592+0.353i\;, \\
  h_{\mu 1} = 0.351+1.17i\;,\quad 
  h_{\mu 2} = 0.329-0.098i\;,\quad 
  h_{\mu 3} = 1.29+0.045i\;,\\
   h_{\tau 1} = 3.40\times 10^{-6}\;,\quad 
   h_{\tau 2} = 1.56\times 10^{-3}\;,\quad 
   h_{\tau 3} = 1.61\;.\qquad
\end{eqnarray}
This set of parameters will then lead to an asymmetry of $\eta_B \simeq 1.5\times 10^{-10}$, which means baryogenesis is marginally successful in this case.\footnote{We have tested many other sets of parameters for this, and the example shown here is one of those more promising ones.}

\subsection{Quasi-degenerate case}

Next, we address the case with quasi-degenerate mass spectrum depicted in (\ref{eqn:QD_spectrum}). Again, in order to ensure that the sizes of the Yukawas are within the perturbative regime, we pick $\lambda_{\phi\chi}' = 1$. This gives
\beqn{eqn:M_nu_real_QD}
 \frac{5760\pi^2}{7\lambda_{\phi\chi}' \langle\phi^0\rangle^2}\, 
  M_\nu^\text{QD}
 \simeq
 \ththMat{429.65}{1.8158}{-1.3998}
         {1.8158}{439.38}{-9.7904}
         {-1.3998}{-9.7904}{439.44} \times 10^{-22} \text{ eV}^{-1}\;.
\eeqn
However, given this restriction on $\lambda_{\phi\chi}'$, it turns out that the resultant baryon asymmetry here is at best around $6\times 10^{-11}$. For example, if we impose
\beqn{eqn:QD_Ks}
 K_{z\tau} \simeq 1.2 \;, \qquad
 K_{x\tau} \simeq 0.01 \;,
\eeqn
and select
\begin{eqnarray}
 h_{e1} = 3.25-1.91i\;,\quad 
 h_{e2} = 0.541+0.895i\;,\quad 
 h_{e3} = -0.090-0.128i\;, \\
  h_{\mu 1} = 1.972+3.23i\;,\quad 
  h_{\mu 2} = -0.916+0.543i\;,\quad 
  h_{\mu 3} = 0.040-0.078i\;,\\
   h_{\tau 1} = 3.40\times 10^{-6}\;,\quad 
   h_{\tau 2} = 1.21\times 10^{-3}\;,\quad 
   h_{\tau 3} = 4.06\;.\qquad
\end{eqnarray}
Then, the final baryon asymmetry is worked out to be $\eta_B \simeq 4.3\times 10^{-11}$, which is about an order of magnitude below the expected value. As a result, this case is actually the least favored of the three within the specific scenario of leptogenesis discussed here.\\

Overall, it appears that a normal hierarchical light neutrino mass spectrum is the most natural choice. This is because during our parameter space search for successful leptogenesis (given the theoretical errors), we have found more leeway when fitting the normal hierarchical case than the other two. However, this does not automatically imply that the inverted case is disfavored (or the quasi-degenerate case is ruled out) in general since we have assumed a very specific exotic mass spectrum of (\ref{eqn:Mass_Choices}). Furthermore, our leptogenesis analysis in Sec.~\ref{sec:lepto} was based on a hierarchical mass spectrum with the special mass relations given in (\ref{eqn:mass_scale_set}), hence, there is a potentially larger parameter space that is viable for our model than suggested by our specific scenario shown in this paper. 

Finally, we point out that when compared with standard leptogenesis, this model contains one extra type of field, namely the new 6-plet scalar $\chi$, which then effectively gives rise to an additional tunable parameter ($\lambda_{\phi\chi}'$) when fitting the leptogenesis and neutrino parameters. But this increase in parameters and fields (with no new symmetries imposed) resulted in a model that also provides a DM candidate, something standard seesaw models do not automatically allow. Thus, in a sense, this model is almost on a similar footing with standard seesaw as far as new input parameters beyond the SM versus new predictions that it generates. As a result, we believe it is a very interesting alternative solution for baryogenesis, dark matter and neutrino masses.

\section{Conclusion}

In this work, we have analyzed an exotic multiplet model where three RH fermion 5-plets and one scalar 6-plet with quantum numbers $(1,5,0)$ and $(1,6,-1/2)$ respectively are added to the minimal SM. Such selection of new particles is motivated by the ``minimal dark matter'' idea \cite{MDM} where the DM candidate can be made stable in the relevant cosmological timescale without any extra symmetries (local or global) being imposed. For our particular setup where the lightest 5-plet particle ($N_1^0$) is the DM, the SM gauge symmetries automatically forbid any unwanted interaction terms as long as the scalar potential is such that 6-plet $\chi$ does not develop a nonzero VEV. 
The observed relic density  then demands that such DM candidate must have a mass of around 10 TeV.

With this in mind, we have subsequently demonstrated that, by tuning the mass spectrum of the exotic multiplets in the model, the correct baryon asymmetry may be generated via leptogenesis. Specifically, when the masses are chosen as $M_3 \gg M_2 \gg M_\chi \gg M_1 \simeq 10$~TeV, this model possesses all the essential features from the standard ``$N_2$-leptogenesis'' scenario (with flavor effects). So, it is quite natural for it to accommodate the requirements of successful baryogenesis.

Moreover, light neutrino masses for all types of hierarchy schemes can be generated radiatively via one-loop diagrams containing the multiplet particles. By adjusting the elements of the Yukawa matrix, we have shown with concrete examples that all experimental neutrino mass and mixing parameters can be fitted. In addition, for most cases (ie. hierarchical light neutrino spectra), these choices of parameters are fully compatible with the conditions for $N_2$-flavored leptogenesis, and hence, proving that this relatively modest extension of the SM can indeed simultaneously tackle the problems of baryon asymmetry, dark matter and neutrino masses.

\begin{acknowledgments}
The authors would like to thank Damien~George and Raymond~Volkas for comments on the manuscript. SSCL would like to thank Xiao-Gang~He and Jayne~Thompson for useful discussions. This work is supported in part by the NSC (grant numbers: NSC-100-2112-M-006-014-MY3 and NSC-100-2811-M-006-019) and in part by the NCTS of Taiwan.
\end{acknowledgments}

\appendix

\section{Selected Clebsch-Gordon coefficients for SU(2)}\label{App:CG_coef}

For the convenience of the reader, we have collected together some of the Clebsch-Gordon coefficients for contracting the various SU(2) multiplets which appeared in this paper. 

In the following, the SU(2) $n$-plets are denoted by $\underline{n}_x = (x_1, x_2, \ldots, x_n)^T$, where the convention is to order the elements according to their isospin values such that  $m_{x_1} = m_{x_2} +1 = m_{x_3} +2 = \cdots = m_{x_n} +n-1$. Note that $n^*$-plets are related to their $n$-plets counterparts via unitary transformation\footnote{The same unitary transformation that takes $\tau_a \rightarrow -\tau_a^*$, where $\tau_a$'s are the generators of SU(2).}
\beqn{app:ut_nplets}
 \underline{n}_x^* \equiv U_{n\times n} 
 \begin{pmatrix}
   x_1^*\\
   x_2^*\\
   \vdots\\
   x_n^*
 \end{pmatrix}\;.
\eeqn
Examples of these unitary transformation matrices for $n=2,6$ are included below:
\beqn{app:itautwo}
 U_{2\times 2} = \twtwMat{0}{1}{-1}{0}\;,\quad
 U_{6\times 6} =  
    \begin{pmatrix}
    0&0&0&0&0&1\\
    0&0&0&0&-1&0\\
    0&0&0&1&0&0\\
    0&0&-1&0&0&0\\
    0&1&0&0&0&0\\
    -1&0&0&0&0&0
     \end{pmatrix}\;.
\eeqn
Clebsch-Gordon expansions for $\twoR \otimes \twoR = \oneR \oplus \thrR$:
\beqn{app:two_1and3}
 (\twoR_a \otimes \twoR_b)_{\oneR} = \frac{a_1 b_2}{\sqrt{2}}- 
            \frac{a_2 b_1}{\sqrt{2}}\;,
 \qquad
  (\twoR_a \otimes \twoR_b)_{\thrR} = 
     \begin{pmatrix}
           a_1 b_1\\
           \frac{a_1 b_2}{\sqrt{2}}+ \frac{a_2 b_1}{\sqrt{2}}\\
           a_2 b_2
      \end{pmatrix}\;.
\eeqn
For $\fivR \otimes \fivR = \oneR \oplus \thrR  \oplus \fivR  \oplus \sevR  \oplus \ninR$:
\beqn{app:fiv_1and3}
 (\fivR_a \otimes \fivR_b)_{\oneR} =
   \frac{a_1 b_5}{\sqrt{5}}
   -\frac{a_2 b_4}{\sqrt{5}}
   +\frac{a_3 b_3}{\sqrt{5}}
   -\frac{a_4 b_2}{\sqrt{5}}
   +\frac{a_5 b_1}{\sqrt{5}}\;,
 \quad
  (\fivR_a \otimes \fivR_b)_{\thrR} = 
   \begin{pmatrix}
      \frac{a_1 b_4}{\sqrt{5}}
      -\frac{a_2 b_3\sqrt{3}}{\sqrt{10}}
      +\frac{a_3 b_2\sqrt{3}}{\sqrt{10}}
      -\frac{a_4 b_1}{\sqrt{5}}\\
      \frac{a_1 b_5\sqrt{2}}{\sqrt{5}}
      -\frac{a_2 b_4}{\sqrt{10}}
      +\frac{a_4 b_2}{\sqrt{10}}
      -\frac{a_5 b_1\sqrt{2}}{\sqrt{5}}\\
      \frac{a_2 b_5}{\sqrt{5}}
      -\frac{a_3 b_4\sqrt{3}}{\sqrt{10}}
      +\frac{a_4 b_3\sqrt{3}}{\sqrt{10}}
      -\frac{a_5 b_2}{\sqrt{5}}
      \end{pmatrix}\;.
\eeqn
$(\fivR_a \otimes \fivR_b)_{\fivR\,,\,\sevR\,,\,\ninR}$ are not shown.\\

For $\sixR \otimes \sixR = \oneR \oplus \thrR  \oplus \fivR  \oplus \sevR  \oplus \ninR \oplus \eleR$:
\begin{align}
 (\sixR_a \otimes \sixR_b)_{\oneR} &=
   \frac{a_1 b_6}{\sqrt{6}}
   -\frac{a_2 b_5}{\sqrt{6}}
   +\frac{a_3 b_4}{\sqrt{6}}
   -\frac{a_4 b_3}{\sqrt{6}}
   +\frac{a_5 b_2}{\sqrt{6}}
   -\frac{a_6 b_1}{\sqrt{6}}\;,
   \\
  (\sixR_a \otimes \sixR_b)_{\thrR} &= 
   \begin{pmatrix}
      \frac{a_1 b_5}{\sqrt{7}}
      -\frac{a_2 b_4\sqrt{8}}{\sqrt{35}}
      +\frac{3 a_3 b_3}{\sqrt{35}}
      -\frac{a_4 b_2\sqrt{8}}{\sqrt{35}}
      +\frac{a_5 b_1}{\sqrt{7}}\\
      \frac{a_1 b_6\sqrt{5}}{\sqrt{14}}
      -\frac{3 a_2 b_5}{\sqrt{70}}
      +\frac{a_3 b_4}{\sqrt{70}}
      +\frac{a_4 b_3}{\sqrt{70}}
      -\frac{3 a_5 b_2}{\sqrt{70}}
      +\frac{a_6 b_1\sqrt{5}}{\sqrt{14}}\\
      \frac{a_2 b_6}{\sqrt{7}}
      -\frac{a_3 b_5\sqrt{8}}{\sqrt{35}}
      +\frac{3 a_4 b_4}{\sqrt{35}}
      -\frac{a_5 b_3\sqrt{8}}{\sqrt{35}}
      +\frac{a_6 b_2}{\sqrt{7}}
      \end{pmatrix}\;,\\
      (\sixR_a \otimes \sixR_b)_{\fivR} &= 
   \begin{pmatrix}
      \frac{a_1 b_4\sqrt{5}}{\sqrt{28}}
      -\frac{3 a_2 b_3}{\sqrt{28}}
      +\frac{3 a_3 b_2}{\sqrt{28}}
      -\frac{a_4 b_1\sqrt{5}}{\sqrt{28}}\\
       \frac{a_1 b_5\sqrt{5}}{\sqrt{14}}
      -\frac{a_2 b_4}{\sqrt{7}}
      +\frac{a_4 b_2}{\sqrt{7}}
      -\frac{a_5 b_1\sqrt{5}}{\sqrt{14}}\\
      \frac{5 a_1 b_6}{\sqrt{84}}
      +\frac{a_2 b_5}{\sqrt{84}}
      -\frac{2 a_3 b_4}{\sqrt{21}}
      +\frac{2 a_4 b_3}{\sqrt{21}}
      -\frac{a_5 b_2}{\sqrt{84}}
      -\frac{5 a_6 b_1}{\sqrt{84}}\\
       \frac{a_2 b_6\sqrt{5}}{\sqrt{14}}
      -\frac{a_3 b_5}{\sqrt{7}}
      +\frac{a_5 b_3}{\sqrt{7}}
      -\frac{a_6 b_2\sqrt{5}}{\sqrt{14}}\\
      \frac{a_3 b_6\sqrt{5}}{\sqrt{28}}
      -\frac{3 a_4 b_5}{\sqrt{28}}
      +\frac{3 a_5 b_4}{\sqrt{28}}
      -\frac{a_6 b_3\sqrt{5}}{\sqrt{28}}
   \end{pmatrix}      
      \;.
\end{align}
$(\sixR_a \otimes \sixR_b)_{\sevR\,,\,\ninR\,,\,\eleR}$ are not shown. And
\beqn{app:77_1}
(\sevR_a \otimes \sevR_b)_{\oneR} =
    \frac{a_1 b_7}{\sqrt{7}}
   -\frac{a_2 b_6}{\sqrt{7}}
   +\frac{a_3 b_5}{\sqrt{7}}
   -\frac{a_4 b_4}{\sqrt{7}}
   +\frac{a_5 b_3}{\sqrt{7}}
   -\frac{a_6 b_2}{\sqrt{7}}
   +\frac{a_7 b_1}{\sqrt{7}}\;.
\eeqn

For $\twoR \otimes \fivR = \fouR \oplus \sixR$:
\beqn{app:twofive_6}
 (\twoR_a \otimes \fivR_b)_{\sixR} = 
    \begin{pmatrix}
    a_1 b_1\\
    \frac{2 a_1 b_2\sqrt{2}}{\sqrt{5}}+\frac{a_2 b_1}{\sqrt{5}}\\
    \frac{2 a_1 b_3\sqrt{3}}{\sqrt{5}}+\frac{a_2 b_2\sqrt{2}}{\sqrt{5}}\\
    \frac{2 a_1 b_4\sqrt{2}}{\sqrt{5}}+\frac{a_2 b_3\sqrt{3}}{\sqrt{5}}\\    
    \frac{a_1 b_5\sqrt{2}}{\sqrt{5}}+\frac{2a_2 b_4}{\sqrt{5}}\\
    a_2 b_5
   \end{pmatrix}  \;.
\eeqn
$(\twoR_a \otimes \fivR_b)_{\fouR}$ is not shown.\\

For $\twoR \otimes \sixR = \fivR \oplus \sevR$:
\beqn{app:twosix_57}
 (\twoR_a \otimes \sixR_b)_{\fivR} = 
    \begin{pmatrix}
    \frac{a_1 b_2}{\sqrt{6}}-\frac{a_2 b_1\sqrt{5}}{\sqrt{6}}\\
    \frac{a_1 b_3}{\sqrt{3}}-\frac{a_2 b_2\sqrt{2}}{\sqrt{3}}\\
    \frac{a_1 b_4}{\sqrt{2}}-\frac{a_2 b_3}{\sqrt{2}}\\
    \frac{a_1 b_5\sqrt{2}}{\sqrt{3}}-\frac{a_2 b_4}{\sqrt{3}}\\
    \frac{a_1 b_6\sqrt{5}}{\sqrt{6}}-\frac{a_2 b_5}{\sqrt{6}}    
   \end{pmatrix}  \;,
\quad
    (\twoR_a \otimes \sixR_b)_{\sevR} = 
    \begin{pmatrix}
    a_1 b_1\\
    \frac{a_1 b_2\sqrt{5}}{\sqrt{6}}+\frac{a_2 b_1}{\sqrt{6}}\\
    \frac{a_1 b_3\sqrt{2}}{\sqrt{3}}+\frac{a_2 b_2}{\sqrt{3}}\\
    \frac{a_1 b_4}{\sqrt{2}}+\frac{a_2 b_3}{\sqrt{2}}\\
    \frac{a_1 b_5}{\sqrt{3}}+\frac{a_2 b_4\sqrt{2}}{\sqrt{3}}\\
    \frac{a_1 b_6}{\sqrt{6}}+\frac{a_2 b_5\sqrt{5}}{\sqrt{6}}\\   
    a_2 b_6
   \end{pmatrix}  \;.
\eeqn

\section{Mass relations for the 6-plet scalar}\label{app:chi_mass}

In this section, we list the tree-level mass relations for all components of 6-plet scalar $\chi$ in terms of the parameters in potential $V_S$ and where $\langle\phi\rangle = (0,u)^T$ with $u \neq 0$.
\begin{align}
 M^2_{\chi^{---}} &:\quad \frac{\mu_\chi^2}{\sqrt{6}}
   +\frac{7-\sqrt{35}}{14\sqrt{3}}\, u^2 \lambda_{\phi\chi 1}
   +\frac{1}{\sqrt{7}}\, u^2 \lambda_{\phi\chi 2}
   -\frac{7\sqrt{5}+\sqrt{7}}{42}\, u^2 \lambda_{\phi\chi 3}
   \;,
   \\
 M^2_{\rm{Re}[\chi^{0}]} &:\quad   \frac{\mu_\chi^2}{\sqrt{6}}
   +\frac{35+\sqrt{35}}{70\sqrt{3}} \, u^2\lambda_{\phi\chi 1}
   +\frac{7\sqrt{5}+5\sqrt{7}}{70}\,u^2\lambda_{\phi\chi 2}
   -\frac{7\sqrt{5}+10\sqrt{7}}{105}\,u^2\lambda_{\phi\chi 3}
   +\frac{7\sqrt{5}-5\sqrt{7}+2\sqrt{105}}{35}\,u^2{\lambda}_{\phi\chi}'
   \;,
 \\
 M^2_{\rm{Im}[\chi^{0}]} &:\quad   \frac{\mu_\chi^2}{\sqrt{6}}
 +\frac{35+\sqrt{35}}{70\sqrt{3}} \, u^2\lambda_{\phi\chi 1}
   +\frac{7\sqrt{5}+5\sqrt{7}}{70}\,u^2\lambda_{\phi\chi 2}
   -\frac{7\sqrt{5}+10\sqrt{7}}{105}\,u^2\lambda_{\phi\chi 3}
   -\frac{7\sqrt{5}-5\sqrt{7}+2\sqrt{105}}{35}\,u^2{\lambda}_{\phi\chi}'
   \;,
\end{align}
and for the components $\chi^{++},\chi^{+},\chi^{-},\chi^{--}$, the mass matrix is non-diagonal:
\beqn{eqn:chi_massmatrix}
\left((\chi^{++})^\dagger,(\chi^{+})^\dagger ,\chi^{-},\chi^{--}\right)
 \begin{pmatrix}
  C_1& 0  & 0  & C_6\\
  0  & C_2& C_5& 0\\
  0  & C_5& C_3& 0\\
  C_6& 0  & 0  & C_4\\
 \end{pmatrix}
  \begin{pmatrix}
  \chi^{++}\\
  \chi^{+}\\
  (\chi^{-})^\dagger\\
  (\chi^{--})^\dagger
 \end{pmatrix}
 \;,
\eeqn
where 
\begin{align}
C_1 &= \frac{\mu_\chi^2}{\sqrt{6}}
   + \frac{7+\sqrt{35}}{14\sqrt{3}}u^2\lambda_{\phi\chi 1}
   + \frac{7\sqrt{5}+\sqrt{7}}{42}u^2\lambda_{\phi\chi 2}
   -\frac{1}{\sqrt{7}}u^2\lambda_{\phi\chi 3}\;, \\
C_2 &= \frac{\mu_\chi^2}{\sqrt{6}}
   + \frac{35+3\sqrt{35}}{70\sqrt{3}}u^2\lambda_{\phi\chi 1}
   + \frac{14\sqrt{5}+5\sqrt{7}}{105}u^2\lambda_{\phi\chi 2}
   -\frac{7\sqrt{5}+25\sqrt{7}}{210}u^2\lambda_{\phi\chi 3}\;, \\
 C_3 &=\frac{\mu_\chi^2}{\sqrt{6}}
   + \frac{35-\sqrt{35}}{70\sqrt{3}}u^2\lambda_{\phi\chi 1}
   + \frac{7\sqrt{5}+10\sqrt{7}}{105}u^2\lambda_{\phi\chi 2}
   -\frac{7\sqrt{5}+5\sqrt{7}}{70}u^2\lambda_{\phi\chi 3}\;,\\
C_4 &=\frac{\mu_\chi^2}{\sqrt{6}}
   + \frac{35-3\sqrt{35}}{70\sqrt{3}}u^2\lambda_{\phi\chi 1}
   + \frac{7\sqrt{5}+25\sqrt{7}}{210}u^2\lambda_{\phi\chi 2}
   -\frac{14\sqrt{5}+5\sqrt{7}}{105}u^2\lambda_{\phi\chi 3}\;, \\
C_5 &= \frac{2(5\sqrt{14}-7\sqrt{10}- 2\sqrt{210})}{105} u^2{\lambda}_{\phi\chi}'\;,\\
C_6 &= \frac{7+2\sqrt{21}- \sqrt{35}}{21} u^2{\lambda}_{\phi\chi}' \;.
\end{align}
Note that only the quartic coupling $\lambda_{\phi\chi}'$ contributes to the mixing terms $C_5$ and $C_6$.


\end{document}